\documentclass[12pt]{article}
\usepackage{graphicx}
\usepackage[left=1in, right=1in]{geometry}

\title{VIP 2: Experimental tests of the Pauli Exclusion Principle for electrons}

\author{A. Pichler$^{a}$, \vspace{-2ex} S. Bartalucci$^{b}$, \vspace{-2ex} M. Bazzi$^{b}$, \vspace{-2ex} S. Bertolucci$^{c}$, \vspace{-2ex} C. Berucci$^{a,b}$, \and M. Bragadireanu$^{b,d}$, \vspace{-2ex}  M. Cargnelli$^{a}$, \vspace{-2ex} A. Clozza$^{b}$, \vspace{-2ex} C. Curceanu$^{b,d,j}$, \vspace{-2ex} L. De Paolis$^{b}$, \and  S. Di Matteo$^{e}$, \vspace{-2ex} A. D'Uffizi$^{b}$, \vspace{-2ex} J.-P. Egger$^{f}$, \vspace{-2ex} C. Guaraldo$^{b}$, \vspace{-2ex} M. Iliescu$^{b}$, \and T. Ishiwatari$^{a}$, \vspace{-2ex} M. Laubenstein$^{g}$, \vspace{-2ex} J. Marton$^{a}$, \vspace{-2ex} E. Milotti$^{h}$, \vspace{-2ex} D. Pietreanu$^{b,d}$, \and K. Piscicchia$^{b,j}$, \vspace{-2ex} T. Ponta$^{b}$, \vspace{-2ex} E. Sbardella$^{b}$, \vspace{-2ex} A. Scordo$^{b}$, \vspace{-2ex} H. Shi$^{b}$, \vspace{-2ex} D. Sirghi$^{d}$, \and F. Sirghi$^{b,d}$, \vspace{-2ex} L. Sperandio$^{b}$, \vspace{-2ex} O. Vazquez-Doce$^{i}$, \vspace{-2ex} E. Widmann$^{a}$, \vspace{-2ex} J. Zmeskal$^{a}$}
\date{}
\begin{document}
\maketitle
\noindent \llap{$^{a}$}Stefan Meyer Institute for subatomic physics, Boltzmanngasse 3, 1090 Vienna, Austria \\
\llap{$^{b}$}INFN, Laboratori Nazionali di Frascati, C.P. 13, Via E. Fermi 40, I-00044 Frascati(Roma), Italy\\
\llap{$^{c}$}CERN, CH-1211, Geneva 23, Switzerland\\
\llap{$^{d}$}IFIN-HH, Institutul National pentru Fizica si Inginerie Nucleara Horia Hulubbei, Reactorului 30, Magurele, Romania\\
\llap{$^{e}$}Institut de Physique UMR CNRS-UR1 6251, Universit\'e de Rennes, F-35042 Rennes, France\\
\llap{$^{f}$}Institut de Physique, Universit\'{e} de Neuch\^{a}tel, 1 rue A.-L. Breguet, CH-2000 Neuch\^{a}tel, Switzerland\\
\llap{$^{g}$}INFN, Laboratori Nazionali del Gran Sasso, I-67010 Assergi (AQ), Italy\\
\llap{$^{h}$}Dipartimento di Fisica, Universit\`{a} di Trieste and INFN-Sezione di Trieste, Via Valerio, 2, I-34127 Trieste, Italy\\
\llap{$^{i}$}Excellence Cluster Universe, Technische Universit\"at M\"unchen, Boltzmannstrasse 2, D-85748 Garching, Germany\\
\llap{$^{j}$}Museo Storico della Fisica e Centro Studi e Ricerche Enrico Fermi, Piazza del Viminale 1, 00183 Roma, Ital

\begin{abstract}
 The Pauli Exclusion Principle (PEP) was famously discovered in 1925 by the austrian physicist Wolfgang Pauli. Since then, it underwent several experimental tests. Starting in 2006, the VIP (Violation of the Pauli Principle) experiment looked for 2p to 1s X-ray transitions in copper, where 2 electrons are present in the 1s state before the transition happens. These transitions violate the PEP, and the lack of detection of the corresponding X-ray photons lead to a preliminary upper limit for the violation of the PEP of $4.7 \times 10^{-29}$. The follow-up experiment VIP 2 is currently in the testing phase and will be transported to its final destination, the underground laboratory of Gran Sasso in Italy, in autumn 2015. Several improvements compared to its predecessor like the use of new X-ray detectors and active shielding from background gives rise to a goal for the improvement of the upper limit of the probability for the violation of the Pauli Exclusion Principle of 2 orders of magnitude.
\end{abstract}

%\FullConference{The European Physical Society Conference on High Energy Physics\\
%		22--29 July 2015\\
	%	Vienna, Austria}
%\PACS{03.65.-w, 07.85Fv,32.30.Rj}

\section{Introduction}
\label{sec:Introduction}
The Pauli Exclusion Principle states that two fermions can not be in the same quantum state \cite{Pauli}. It is, according to current knowledge, a consequence of the Spin Statistics connection. Wolfgang Pauli was awarded the Nobel Prize in physics in 1945 for its discovery. It is a fundamental principle in nature, mainly where many-fermion systems are concerned. Therefore, it needs to be tested thoroughly. At present, an intuitive explanation for the PEP is still missing \cite{Feynman}.

To the best of our current understanding, there are only 2 spin-separated classes of particles: fermions with half-integer spin and bosons with integer spin. The PEP is only fulfilled for fermions. Several experiments have tested the Pauli Exclusion Principle in the past. Some of them have put very stringent limits on the probability for the violation of the PEP in stable systems. These results have for example been put forward by the DAMA collaboration \cite{DAMA}. This kind of experiments looked for forbidden transitions in stable systems. But those transitions violate the Messiah-Greenberg superselection rule \cite{MG}, which forbids the change of the symmetry state of a system, which is not interacting with its surroundings. To circumvent this rule, the idea was introduced to use ``new'' electrons for testing the Pauli Exclusion Principle.

\section{Experimental Method}
\label{sec:ExpMethod}
\begin{figure}[ht]
 \centering
 \includegraphics[width=0.70\textwidth]{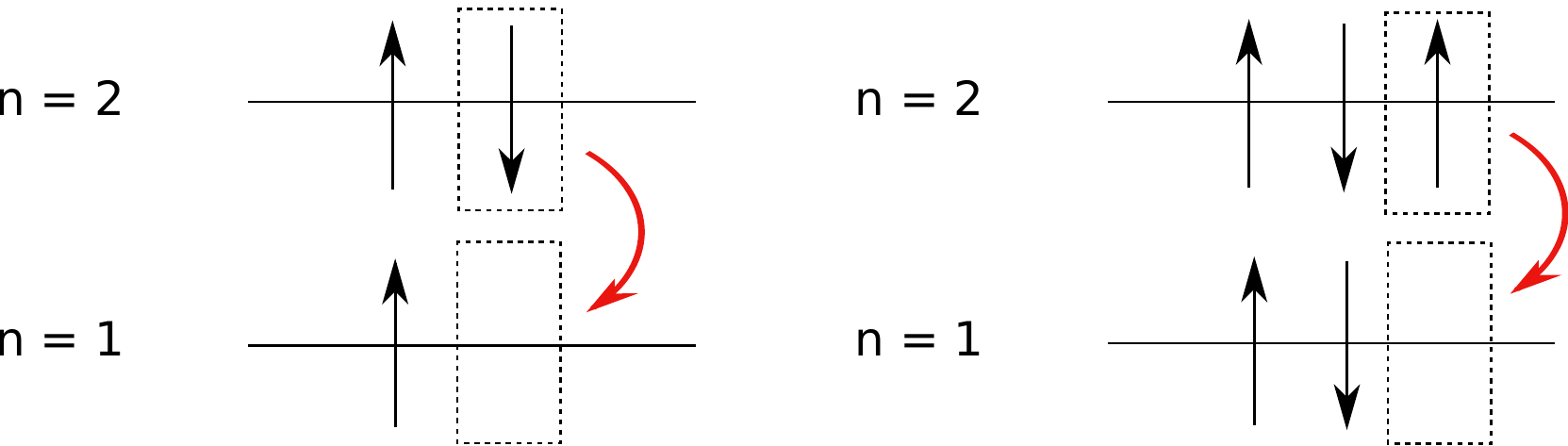}
 % Fig1.png: 1992x637 pixel, 120dpi, 42.17x13.48 cm, bb=0 0 1195 382
 \caption{Normal (allowed) 2p - 1s transition with an energy of 8.05 keV for copper (left) and non-Paulian (forbidden) transition  with an energy of around 7.7 keV for copper (right).}
 \label{fig:transition}
\end{figure}

The idea of the experiments with ``new'' electrons was to introduce these electrons into a target. There they interact with the atoms of the material and have a certain chance to be captured by them. In the course of this process, the absorbed electron forms a new electronic state with the electrons in the atom. This state potentially has a new symmetry, which distinguishes this method from investigations on stable systems. During the cascading process the electrons undergo while they relax into the ground state, they emit radiation, which can be detected. If the PEP is violated in the atom with the newly formed symmetry state, transitions like on the right side of figure \ref{fig:transition} can happen.
The important point of the transition on the right hand side of this figure is, that it has a slightly lower transition energy than the one on the left, due to the enhanced shielding caused by the additional electron in the ground state prior to the transition. An excess of detected X-ray photons in the energy range of around 7.7 keV, where the forbidden transition is expected in copper, would be a hint for the violation of the Pauli Exclusion Principle.

\begin{figure}[ht]
 \centering
 \includegraphics[width=0.60\textwidth]{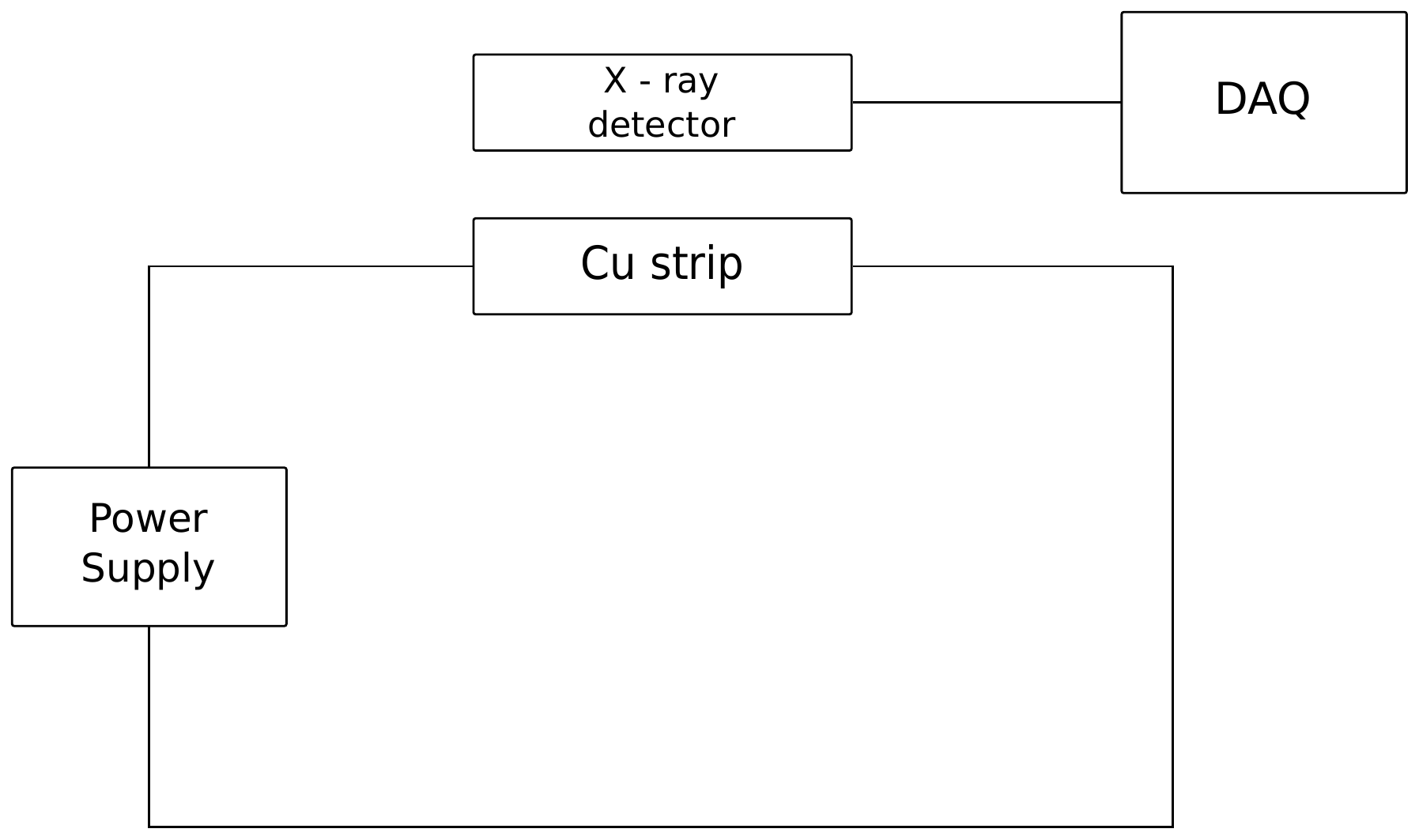}
 % Fig2.pdf: 517x306 pixel, 72dpi, 18.24x10.80 cm, bb=0 0 517 306
 \caption{Scheme of the experiment of Ramberg and Snow.}
 \label{fig:RS_scheme}
\end{figure}

\begin{figure}[ht]
 \centering
 \includegraphics[width=0.9\textwidth]{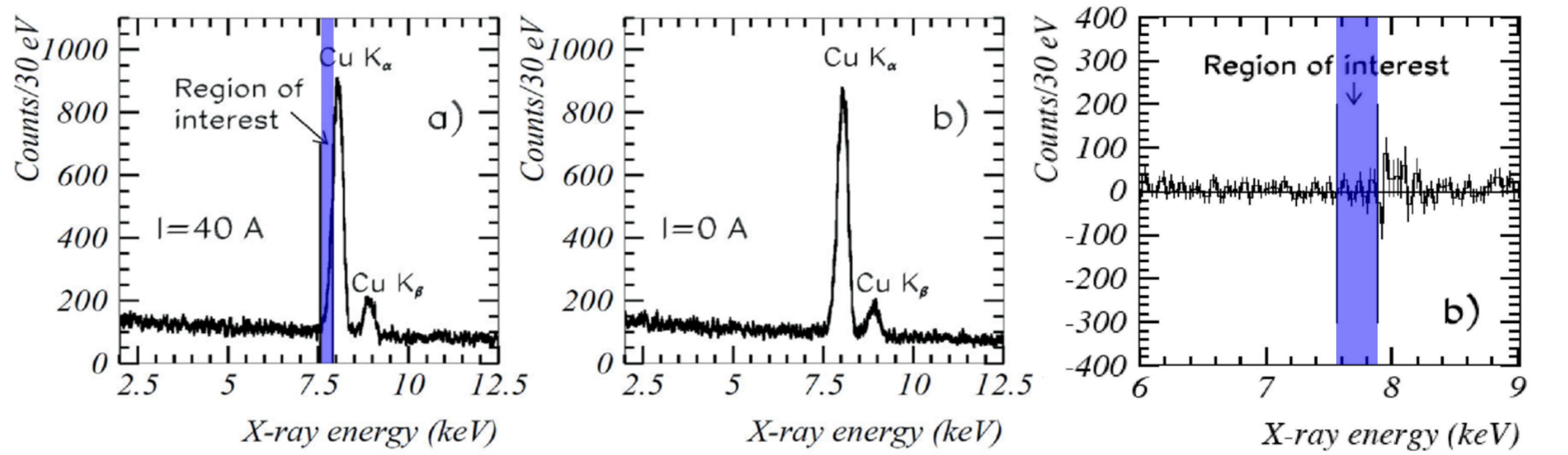}
 % lnf_results.pdf: 1788x518 pixel, 72dpi, 63.08x18.27 cm, bb=0 0 1788 518
 \caption{X-ray spectrum measured with circulating current (left), without current (middle) and the residual plot. The region of interest where Pauli violating K$\alpha$ transitions are expected is marked in blue.}
 \label{fig:lnf_results}
\end{figure}

One of the first experiments that used ``new'' electrons was conducted by Ramberg and Snow \cite{RS}. The scheme of the experiment is shown in figure \ref{fig:RS_scheme}.
They introduced an electric current to a copper strip with the help of a power supply. The current electrons have a chance to be captured by the copper atoms. The experimenters then detected the X-rays, which are emitted during the cascading process of these electrons. The spectrum was measured with and without circulating current. 
The same measurement principle was applied in the VIP experiment. It first took data in late 2005 at the Frascati National Laboratories of INFN in Italy for about 10 days without current and for 10 days with a current of 40 A. In figure \ref{fig:lnf_results}, the results of this data taking period are shown. 
The non-Paulian transitions are only expected in the spectrum with current in the energy ``region of interest'', which is marked in blue. In the residual spectrum on the right hand side of figure \ref{fig:lnf_results} it can be seen that there was no excess found in the spectrum with current compared to the spectrum without current. Thereby an upper bound on the probability for the violation of the PEP could be calculated \cite{VIP2006}:

\begin{equation}
 \frac{\beta^{2}}{2} \leq 4.5 \times 10^{-28}
\end{equation} 
The factor $\frac{\beta^{2}}{2}$ represents the probability, that the Pauli Exclusion Principle is violated in an atom. The VIP experiment furthermore took data in the underground laboratories in Gran Sasso, Italy (LNGS-INFN) until 2010. A preliminary value for $\frac{\beta^{2}}{2}$ was published in \cite{Catalina1, Catalina2}:
\begin{equation}
 \frac{\beta^{2}}{2} \leq 4.7 \times 10^{-29}
\end{equation} 
%
% -----------------------------------------------------

\section{The VIP 2 experiment}
\label{sec:VIP2}
\begin{figure}[h]
 \centering
 \includegraphics[width=0.6\textwidth]{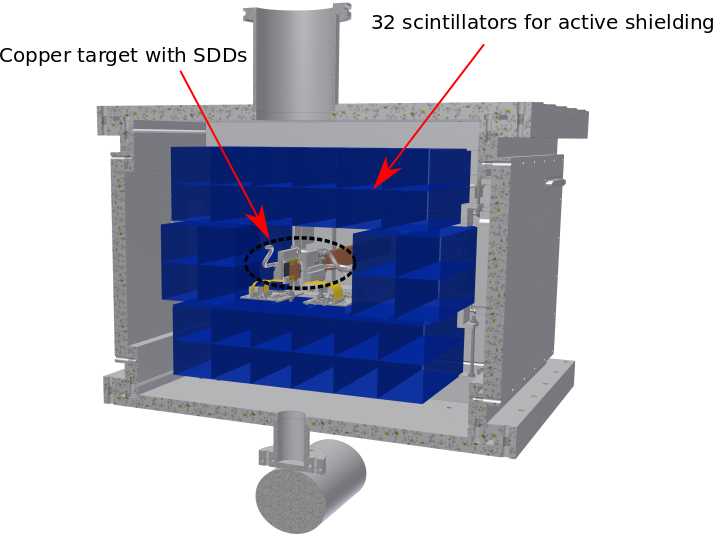}
 % active_shielding.pdf: 571x429 pixel, 72dpi, 20.14x15.13 cm, bb=0 0 571 429
 \caption{View of the VIP 2 setup including the scintillators and the copper target with SDDs.}
 \label{fig:active_shielding}
\end{figure}

The follow-up experiment is currently under preparation at the laboratory of the Stefan Meyer Institute in Vienna. Compared to the VIP experiment, one of the major upgrades is the use of Silicon Drift Detectors (SDDs) as X-ray detectors, whereas the predecessor experiment was using Charge-Coupled Devices (CCDs). On the one hand, SDDs have a better energy resolution, with around 150 eV (FWHM) at 6 keV, compared to around 320 eV for CCDs. On the other hand, they have a time resolution of around 400 ns (FWHM), which enables the possibility of an active shielding of background events, which are caused by cosmic rays. 
For the purpose of active shielding, 32 scintillators read out by Silicon Photomultipliers are arranged around the copper target. All signals recorded by the SDDs, which are in coincidence with signals recorded by the scintillators, can be marked as background events, because both signals are caused by the same cosmic ray. A schematic drawing of the setup is shown in figure \ref{fig:active_shielding}.
Further improvements of the current setup compared to the VIP setup are shown in table \ref{tab:VIP2_improvement}.
\begin{table}
\centering
\caption{List of improvements of VIP 2 compared to VIP \cite{VIP_proposal}.}
\label{tab:VIP2_improvement}
%   \begin{figure}[htbp]
	
		\includegraphics[width=0.7\textwidth]{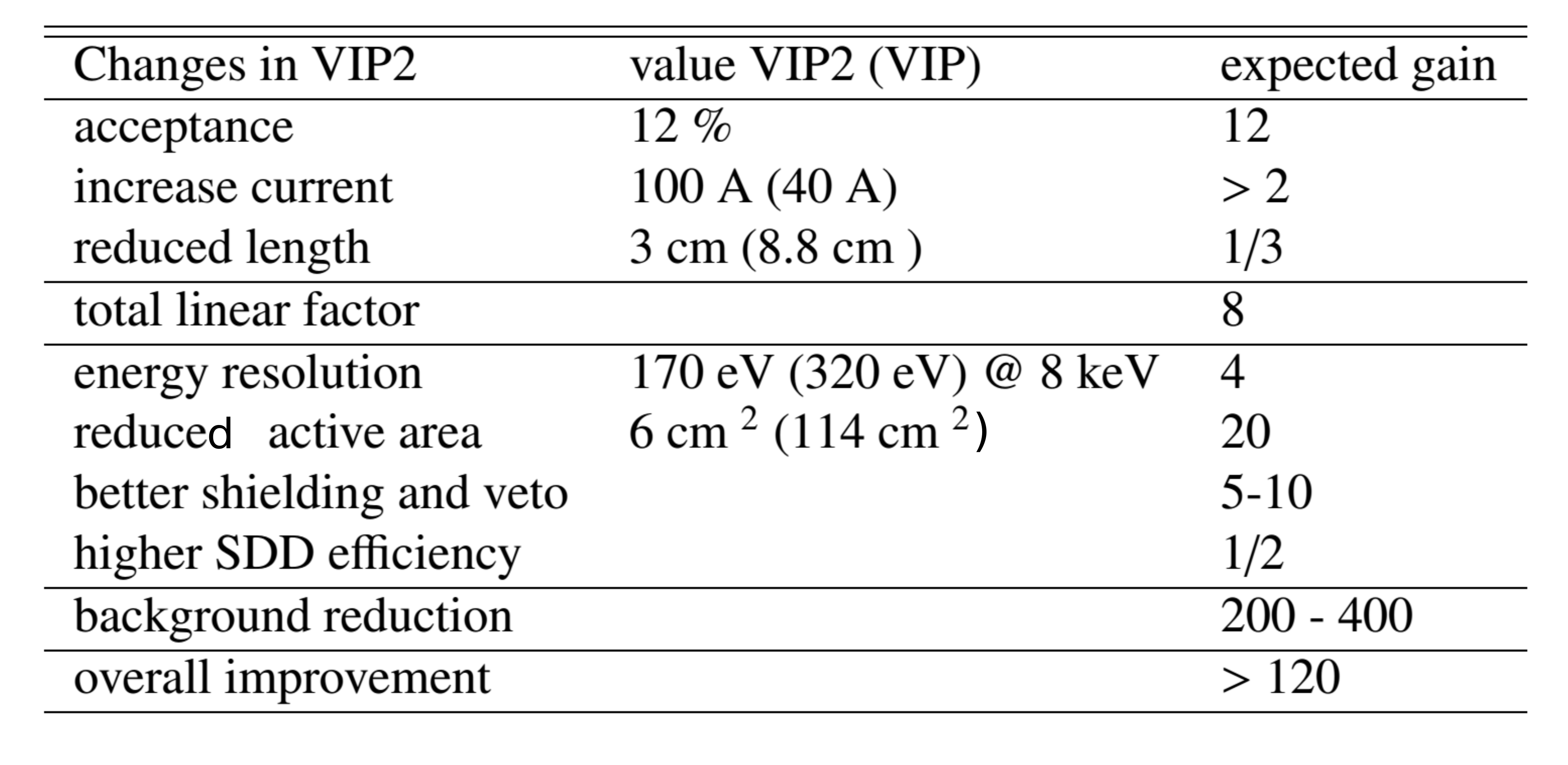}
%	\caption{List of improvements of VIP 2 compared to VIP \cite{VIP_proposal}.}
%	\label{fig:VIP2_improvement}
%  \end{figure}

\end{table} 
Another prominent improvement factor besides the active shielding and the enhanced energy resolution is the increased current through the copper bar, which will result in an improvement of the limit for the violation of the PEP of around $\frac{1}{2}$. All the improvements mentioned above make us confident to aim for a reduction of the limit for the violation of the PEP to around $10^{-31}$. Or else, the violation of the Pauli Exclusion Principle for electrons will be discovered.

\begin{figure}[!htb]
 \centering
 \includegraphics[width=0.6\textwidth]{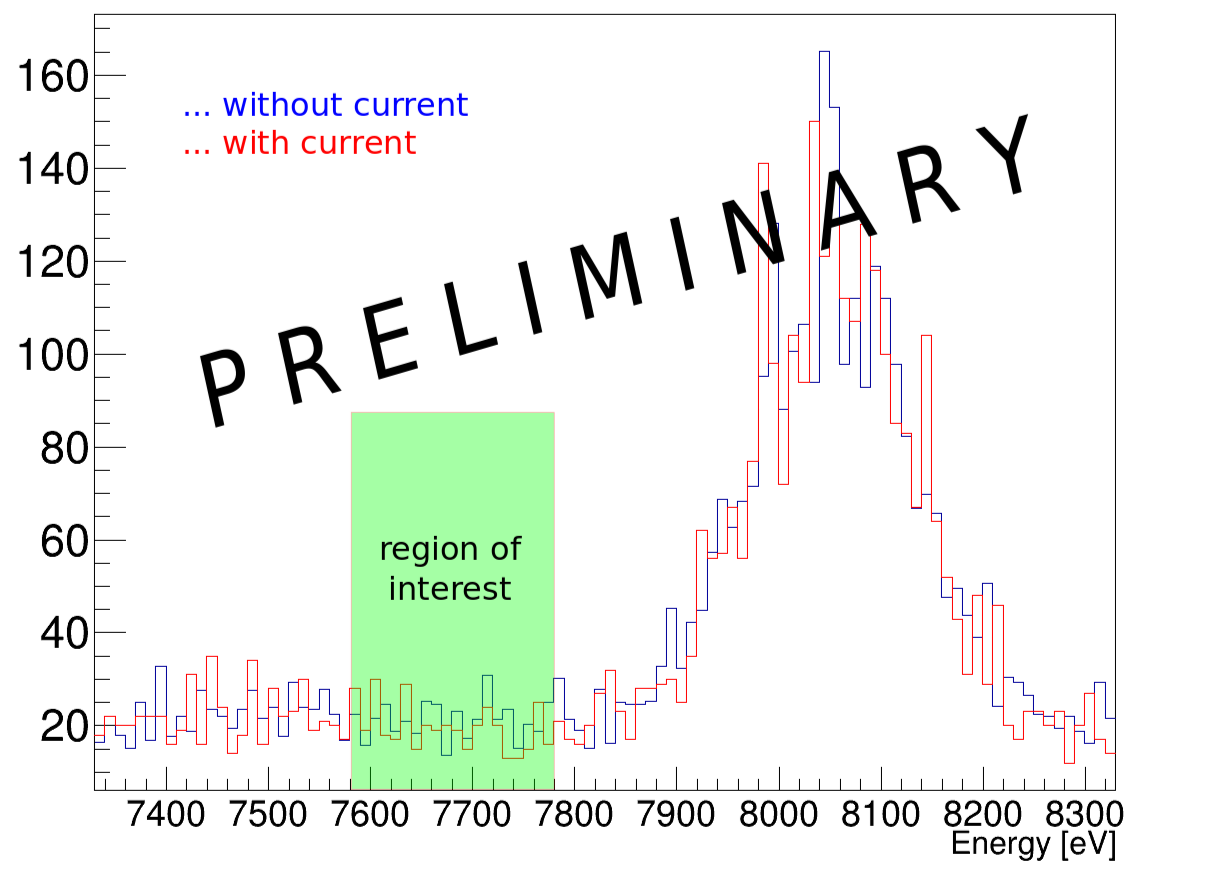}
 % padc_all_wo.png: 1276x971 pixel, 72dpi, 45.01x34.25 cm, bb=0 0 1276 971
 \caption{SDD spectra recorded without current and with a current of 50 A, with the region of interest of the non-Paulian transition marked in green.}
 \label{fig:sdd-spectra}
\end{figure}

First measurements have been conducted in the laboratory at the Stefan Meyer Institute. In figure \ref{fig:sdd-spectra}, a preliminary X-ray spectrum of around 35 hours of data taking without current is shown together with a spectrum of around 35 hours of data taking with a current of 50 A. The spectra are summed up over all 6 used SDDs, which are energy calibrated with the help of a Fe-55 source and a titanium calibration foil.
In the plot, only the part of the spectrum including the copper K$\alpha$-line (around 8 keV) and the region of interest (around 7,7 keV) is shown. The region of interest is marked in green. In this small data set, no excess of counts in this energy region in the spectrum with current could be found compared to the spectrum without current. The calculation of the probability for the violation of the PEP, which corresponds to the parameter $\frac{\beta^{2}}{2}$, from these data is work in progress.
\newpage
\section{Outlook}
\label{sec:outlook}
In autumn 2015, the experiment will be transported to the underground laboratory INFN-LNGS in Gran Sasso, Italy. After implementing the setup and completing stability checks in the new environment, long term data taking of 2-3 years will start.\\ \\
{\small \textbf{Acknowledgement}%
\quad We want to thank H. Schneider, L. Stohwasser and D. St\"uckler from the Stefan Meyer Institute for their important contributions for the design and the construction of the VIP 2 setup and the staff of the INFN-LNGS laboratory for the support during all phases of the experiment. We acknowledge the support from the: HadronPhysics FP6 (506078), HadronPhysics2 FP7 (227431), HadronPhysics3 (283286) projects, EU COST Action 1006 (Fundamental Problems in Quantum Physics), Austrian Science Fund (FWF), which supports the VIP 2 project with the grants P25529-N20 and W1252-N27, Centro Fermi (project: Open problems in quantum mechanics). Furthermore, this paper was made possible through the support of a grant from the John Templeton Foundation (ID 581589). The opinions expressed in this publication are those of the authors and do not necessarily reflect the views of the John Templeton Foundation}

\end{document}